\documentstyle[12pt,epsfig]{article}
\newcommand{\newc}{\newcommand}
\newcommand{\be}[1]{\begin{equation} \label{(#1)}}
\newcommand{\ee}{\end{equation}}
\newcommand{\ba}[1]{\begin{eqnarray} \label{(#1)}}
\newcommand{\ea}{\end{eqnarray}}

\def\pmb#1{\setbox0=\hbox{#1}%
  \kern-.015em\copy0\kern-\wd0
  \kern.03em\copy0\kern-\wd0
  \kern-.015em\raise.0233em\box0 }
\def\Lv{$L\hspace{-0.5em}/ $}

\def\znbb{0\nu\beta\beta}

\def \sw {\sin\!\theta^{}_W }

\def \sw2 {\sin^2\!\theta^{}_W }

\def \s2bt {\sin\!2\beta }

\def \lg  {\langle}
\def \rg  {\rangle}

\def \znbb {0\nu\beta\beta}

\newc{\Rslash}{\not R}
\newc{\superhu}{\hat H_u}
\newc{\superhd}{\hat H_d}
\newc{\superl}{\hat L}      
\newc{\superr}{\hat R}
\newc{\superec}{\hat {e}^c}
\newc{\superq}{\hat Q}
\newc{\superu}{\hat U}
\newc{\superd}{\hat D}
\newc{\superuc}{\hat {u}^c}
\newc{\superdc}{\hat {d}^c}
\newc{\Lsoft}{{\cal L}_{\rm soft}}
\newc{\hc}{{\rm H.c.}}
\newc{\tildeQ}{\widetilde Q}
\newc{\tildeU}{\widetilde U}
\newc{\tildeD}{\widetilde D}
\newc{\tildeL}{\widetilde L}
\newc{\tildeuc}{\widetilde {u}^c}
\newc{\tildedc}{\widetilde {d}^c}
\newc{\tildeec}{\widetilde {e}^c}
\newc{\tildenu}{\widetilde \nu}
\newc{\onehalf}{\frac{1}{2}}
\newc{\gluino}{{\tilde g}}
\newc{\mgluino}{m_{\gluino}}
\newc{\mzero}{m_0}
\newc{\mz}{m_Z}
\newc{\mw}{m_W}
\newc{\alphas}{\alpha_s}
\newc{\tanb}{\tan\beta}
\newc{\hinot}{\widetilde H^0_2}
\newc{\hinob}{\widetilde H^0_1}
\newc{\bino}{\widetilde B^0}
\newc{\wino}{{\widetilde W^0_3}}
\newc{\msl}{m_{\widetilde l}}
\newc{\msel}{m_{\tilde e}}
\newc{\msq}{m_{\tilde q}}
\newc{\msf}{m_{\tilde f}}
\newc{\mev}{{\rm\,MeV}}
\newc{\gev}{{\rm\,GeV}}
\newc{\tev}{{\rm\,TeV}}

\begin{document}

\begin{center}
{\Large \bf The $(\mu^-,e^+)$ conversion in nuclei
mediated by light Majorana neutrinos}

\bigskip {Fedor \v Simkovic$^{a,b}$, Pavol Domin$^b$, Sergey Kovalenko$^c$
\footnote{On leave of absence from the Joint Institute for Nuclear Research, Dubna, Russia}, 
Amand Faessler$^a$}

\bigskip 

{\it 
$^a$Institute f\"ur Theoretische Physik  der Universit\"at
T\"ubingen,
Auf der Morgenstelle 14, D-72076 T\"ubingen, Germany\\[0pt] 
$^b$Department of Nuclear physics, Comenius University,
SK--842 15 Bratislava, Slovakia\\[0pt]
$^c$ Departamento de F\'\i sica, Universidad
T\'ecnica Federico Santa Mar\'\i a, Casilla 110-V, 
Valpara\'\i so, Chile\\[0pt]}

\end{center}
\bigskip

\begin{abstract}
We study the lepton number violating $(\mu^-,e^+)$ conversion in 
nuclei mediated by the exchange of virtual light Majorana neutrinos. 
We found that a previously overlooked imaginary part of this amplitude plays 
an important role. 
The numerical calculation has been made 
for the experimentally interesting $(\mu^-,e^+)$ conversion 
in $^{48}Ti$ using realistic 
renormalized proton-neutron QRPA wave functions. 
We also discuss
the very similar case of the neutrinoless double beta decay
of $^{48}Ca$. The ratio of $(\mu^-,e^+)$ conversion over
the total $\mu^-$ absorption
has been computed taking into account 
the current 
constraints from neutrino oscillation phenomenology. We compare our results  
with the experimental limits as well as with previous theoretical predictions. 
We have found that the  Majorana neutrino mode of $(\mu^-,e^+)$ 
conversion in $^{48}Ti$ 
is  too small to be measurable in the foreseeable future. 
\end{abstract}

\section{Introduction}
\label{sec:level1}

Lepton number (L) conservation is one of the most obscure sides of the standard model(SM) 
not supported by an underlying principle and following from an accidental interplay 
between gauge symmetry and field content. Any deviation from the SM structure may 
introduce L  non-conservation(\Lv).
Over the years the possibility of lepton number non-conservation has been 
attracting a great 
deal of theoretical and experimental efforts since any positive 
experimental \Lv $ $ signal 
would request  physics beyond the SM. 
In addition it would also show, that neutrinos 
are Majorana particles \cite{theorem}. 

Recent neutrino oscillation experiments practically established the presence of
non-zero neutrino masses, a fact that itself points to  physics beyond the SM.
However neutrino oscillations are not sensitive to the nature of neutrino masses: 
they can either be Majorana or Dirac masses leading to the same observables.

The principal question if neutrinos are Majorana or Dirac particles can 
be answered only by studying a lepton number violating processes since $\Delta L =2$
is a generic tag of Majorana neutrinos.
Various lepton-number violating processes have been discussed in the literature
in this respect (for recent review see \cite{DGKS:NANPino}). They offer the possibility of 
probing different entries of the Majorana neutrino mass matrix $M_{ij}^{(\nu)}$. 
Among them there are a few \Lv$ $ nuclear processes having prospects for experimental
searches: neutrinoless double beta decay ($\znbb$), muon to positron $(\mu^-,e^+)$ conversion 
and, probably,  muon to antimuon $(\mu^-,\mu^+)$ conversion \cite{Mohapatra:mu}. They probe 
$M_{ee}^{(\nu)}$, $M_{\mu e}^{(\nu)}$ and $M_{\mu\mu}^{(\nu)}$ matrix elements respectively.  
Currently the most
sensitive experiment intended to distinguish the Majorona nature of neutrinos are those searching 
for neutrinoless $\znbb$-decay \cite{znbb-exp,ejiri,znbb-rev2}             
The nuclear theory side 
\cite{doi,fae98} of this 
process have been significantly improved in the last decade \cite{fae98} 
allowing reliable extraction of fundamental particle physics parameters from experimental data. 
Muon to positron nuclear conversion $(\mu^-,e^+)$ is another \Lv$ $  nuclear process 
with  good experimental prospects. 

The important role of muon as a test particle in the search for 
new physics beyond the standard model has been recognized long time ago.
When negative muons penetrate into matter they can be trapped 
to atomic orbits. Then the bound muon can disappear either 
decaying into an electron and two neutrinos or being captured 
by the nucleus, i.e., due to  ordinary muon capture. These two processes 
conserving both total lepton number and lepton flavors have been well studied both 
theoretically and experimentally. However, there are two 
other not yet  observed  channels of muon capture: 
The muon-electron ($\mu^-, e^-$) and muon-positron ($\mu^-, e^+$)
conversions in nuclei \cite{kamal,vergr,leon,ko94,ko97,mue,siis}:
\begin{eqnarray}
(A,Z) + \mu_b^- &\rightarrow & e^- + (A,Z)^*, \\ \nonumber
(A,Z) + \mu_b^- &\rightarrow & e^+ + (A,Z-2)^*.
\label{eq.1}
\end{eqnarray}
Apparently,  the ($\mu^-, e^+$) and ($\mu^-, e^-$) conversion processes violate 
lepton number L and lepton flavor $L_f$ conservation respectively. 
Additional differences between the $(\mu^-, e^-)$ and 
$(\mu^-, e^+)$ lie on the nuclear physics side. The first process 
can proceed on one nucleon of the participating nucleus while the second process
involves two nucleons as dictated by charge conservation
\cite{vergr,leon}. Note also that the $(\mu^-, e^-)$ conversion 
amplitude is quadratic and $(\mu^-, e^+)$ amplitude linear in the neutrino mass. 
Thus the second process looks more sensitive to the light neutrino masses.
The present experimental limit on  
$(\mu^-,e^+)$ conversion branching ratio in $^{48}$Ti is \cite{doh93,hon96}
\begin{eqnarray}
R_{\mu e^+}(Ti) = \frac{\Gamma(\mu^-+ ^{48}Ti \rightarrow e^+ + ^{40}Ca)}
{\Gamma(\mu^- + ^{48}Ti \rightarrow \nu_{\mu} + ^{48}Sc)}\ <\  4.3\times 10^{-12}. 
\label{eq.2}
\end{eqnarray}

In the present paper we study  
the light Majorana neutrino mechanism for the $\mathbf{(\mu^-, e^+)}$ conversion.
Despite the previous rough estimates \cite{DGKS:NANPino} indicate 
a very small branching ratio for this mode of $\mathbf{(\mu^-, e^+)}$ conversion, 
by far below the experimental bound  (\ref{eq.2}), we undertake a detailed study of 
this mode for several reasons. First, the nuclear theory of $\mathbf{(\mu^-, e^+)}$ 
conversion is not yet sufficiently elaborated, as in the case of $\znbb$-decay, and requires 
further development. Second, 
$\mathbf{(\mu^-, e^+)}$ conversion may receive contribution from other mechanisms 
offered by various models beyond the SM such as the R-parity violating supersymmetric 
models, the leptoquark extensions of the SM etc. Some of these mechanisms may involve 
the light neutrino exchange and, therefore,
from the view point of nuclear structure calculations they resemble the ordinary light 
neutrino mechanism. Thus our present study can be viewed as a first step towards 
a more general description of $\mathbf{(\mu^-, e^+)}$ conversion including 
all the possible mechanisms.

Below we develop a detailed nuclear structure theory for the light 
neutrino exchange mechanism of this process on the basis of the nuclear 
proton-neutron renormalized Quasiparticle Random Phase Approximation (pn-QRPA) 
wave functions  \cite{toi95,simn96}.  We perform a realistic calculation of 
the width of this process for the nuclear target $^{48}Ti$ using limits on neutrino 
masses and mixings from neutrino oscillation phenomenology. 
A comparison with the previous estimations of $R_{\mu e^+}$
will be also presented \cite{doi,leon}. 

The paper is organized as follows. The possible values of Majorana neutrino 
masses and mixings are discussed in sect. 2. The amplitude and width of 
$(\mu^-,e^+)$ conversion are derived in sect. 3. The details of
the calculation for the case of $(\mu^-,e^+)$ conversion in $^{48}Ti$ 
and our results are given in sect. 4.  
In sect. 5 we summarize our conclusions.  

\section{Majorana neutrino mass matrix} 
\label{sec:level2}

The finite masses of neutrinos are tightly related to the problem of  
lepton flavor/number violation. The Dirac, Majorona and Dirac-Majorana neutrino
mass terms in the Lagrangian offer different neutrino mixing schemes and allow
various lepton number/flavor violating processes \cite{bil87,bil98,zubmas}. 
The favored neutrino mixing schemes has to accommodate present 
neutrino phenomenology in a natural way, in particular, to answer the question of 
the smallness of neutrino masses compared to the charged lepton ones.
The most prominent guiding principle in this problem 
is the see-saw mechanism which can be realized in various models beyond the SM.
A generic neutrino mass term is given by the formula 
\begin{equation}
{\cal L}^{D+M}  =
-\sum_{l,l'=e,\mu,\tau} ~[~ \frac{1}{2} ~\overline{(\nu'_{l'L})^c}
~(M^M_L)_{l'l} ~\nu'_{lL}+ \frac{1}{2} ~\overline{\nu'_{l'R}}~
(M^M_R)_{l'l}~ (\nu'_{lR})^c + 
\overline{\nu'_{l'R}} ~(M^D)_{l'l}~ \nu'_{lL}~ ] + h.c.
\label{eq.3}
\end{equation}
The first two terms do not conserve the total lepton number $L$. 
Here, 
$\nu'_L$ and $\nu'_R$ 
are the weak doublet and singlet flavor eigenstates.
The indices L and R refer to the left-handed and right-handed chirality
states, respectively, and the superscript $c$ refers to the operation
of charge conjugation. $M^M_L$ and $M^M_R$ are complex non-diagonal 
symmetrical 3x3 matrices. The flavor neutrino fields are superpositions of six
Majorana fields $\nu_i$ with definite masses $m_i$. 
Yanagida, Gell-Mann, Ramond and Slansky suggested that the elements of  
$M^D$ and $M^M_L$ be comparable with the masses of charged 
leptons and the hypothetical scale of lepton number violation ($ M_{LNV} \approx 10^{12} GeV$),
respectively. Then by diagonalization of  the  Dirac-Majorana mass term one ends up with 
the three very light and three very heavy neutrino eigenstates. 
This is the celebrated see-saw mechanism.
 Enlarging the number
of the right handed neutrino states  $\nu_R$ one can introduce sterile 
light mass eigenstates which may play a certain role in  
the explanation of the 
 neutrino oscillation data including the LSND results.
However the active-sterile neutrino oscillations as a dominant channel 
seems to be disfavored according 
to a recent  Super Kamiokande global analysis \cite{SK-global} and work in T\"ubingen
\cite{haug}.

Sticking to the three neutrino scenario one may try to reconstruct the corresponding mass 
matrix from the neutrino oscillation data. This requires certain  assumptions on 
its structure or additional experimental data.
Solar, atmospheric and LSND neutrino data give information on
the neutrino mass square differences $\Delta m^2_{ij}$ as well as on 
the mixing angles of the unitary matrix $U$ \cite{mix99,viss,barg,smi,haug} 
relating the weak 
$\nu'_{lL}$ and mass $\nu_{iL}$ 
neutrino eigenstates   
\begin{equation}
\nu'_{lL} = \sum_{i=1}^3 ~U^{(\nu)}_{li} ~~\nu_{iL} ~~~
(l=~e,~\mu,~\tau). 
\label{eq.8}
\end{equation}
This information can be used to restore the neutrino mass matrix 
inverting its diagonalization as 
\begin{equation}
{\cal M}^{ph} = U^{(\nu)} \cdot diag(m_1, m_2, m_3) \cdot U^{(\nu)T},
\label{eq.9}
\end{equation}
if additional assumptions about the overall mass scale as well as about 
the CP phases  $\zeta^{(i)}_{CP}$ of the neutrino mass eigenstates 
are made. This matrix can be identified with the $3 \times 3$ Majorana 
mass terms $M^M_L$ in (\ref{eq.3}) and used in various phenomenological 
applications, for instance, in analysis of lepton number violating processes. 
The recent literature contains many sophisticated studies made in this direction
(see, for instance, \cite{Lnv-oscil} and references therein).

The elements of the Majorana mass matrix are related to 
the effective Majorana neutrino masses 
$<m_\nu >_{\alpha\beta}$ ($\alpha,\beta~ =~e,~\mu,~\tau$) as
\begin{equation}
{\cal M}^{th}_{\alpha\beta} \equiv 
<m_\nu >_{\alpha\beta} ~= ~ \sum^{3}_k~ U^{(\nu)}_{\alpha k} ~U^{(\nu)}_{\beta k}
 ~ \zeta^{(k)}_{CP} ~ m_k,
\label{eq.10}
\end{equation}
neglecting mixing with heavy neutral states if they exist in the neutrino mass 
spectrum. 
Amplitudes of lepton number violating processes are proportional to the corresponding
effective neutrino masses 
\cite{DGKS:NANPino,bil87}.
Thus $0\nu\beta\beta$-decay amplitude is proportional to $<m_\nu >_{ee}$, the so called effective  
electron neutrino mass \cite{doi,fae98}.
From the currently most stringent lower limit on
$0\nu\beta\beta$-decay half-life of $^{76}Ge$ 
$T^{0\nu}_{1/2} \ge 1.1\times 10^{25} years$  \cite{znbb-exp} one obtains 
$<m_\nu >_{ee} < 0.62~eV$ \cite{pseu99,nanp}.
 The effective Majorana muon neutrino mass
$<m_\nu >_{\mu\mu}$ is related with
the  light neutrino exchange modes of
muonic analog of $0\nu\beta\beta$-decay \cite{Mohapatra:mu}, semipletonic decay of 
kaon $K^+ \rightarrow \pi^- \mu^+\mu^+$  \cite{LS:2000,grib} etc. 
$<m_\nu >_{\mu e}$ enters the amplitude of the $(\mu^-,e^+)$ conversion 
\cite{vergr,leon}
and of the kaon decay into a muon and a positron 
($K^+ \rightarrow \pi^- \mu^+e^+$). Some other elements of  
${\cal M}^{th}$ are associated with rare B-decays \cite{zuber}.   

In Ref. \cite{haug} the maximal allowed values for the elements of 
${\cal M}^{ph}$ have been deduced from solar, atmospheric, LSND data
and restriction coming  from $0\nu\beta\beta$-decay. The result is 
\begin{eqnarray}
 \left(
 \begin{array}{ccc}
  <m_\nu >_{e e} & <m_\nu >_{e \mu} &<m_\nu >_{e \tau} \\
 <m_\nu >_{\mu e}  & <m_\nu >_{\mu \mu} &<m_\nu >_{\mu \tau} \\
 <m_\nu >_{\tau e}  & <m_\nu >_{\tau \mu} & <m_\nu >_{\tau \tau}
 \end{array}
 \right) 
& \le &
 \left(
 \begin{array}{ccc}
  0.60 & 0.97 & 0.85\\
  0.97 & 0.76 & 0.80\\
  0.85 & 0.80 & 1.17
 \end{array}
 \right) 
\mbox{eV}.
\label{eq.11}
\end{eqnarray}
Thus in our analysis of the light neutrino exchange mode of the
$(\mu^-,e^+)$ conversion process we shall assume 
$|<m_\nu >_{\mu e}| \le 0.97~eV$.

We shall also use a more conservative model independent estimate of 
the effective neutrino mass. Atmospheric and solar neutrino oscillation 
data show up 
$\Delta m^2<<(1 eV)^2$ 
suggesting that all the neutrino mass 
eigenstates are approximately degenerate at the 1 eV scale \cite{barg1}. 
This observation in combination with the tritium beta decay endpoint allows 
one to set upper bounds on masses of all the three neutrinos \cite{barg1}
$m_{e,\mu,\tau}\leq 3$eV. Thus in the three neutrino scenario one derives
$ |\lg m_{\nu}\rg_{ij}|  \leq 9\mbox{eV}$ for $i,j = e, \mu, \tau$ 
\cite{DGKS:NANPino,grib}.

\section{The $\mathbf{(\mu^-, e^+)}$ conversion mediated by light neutrinos}
\label{sec:level3}

The process of $\mathbf{(\mu^-, e^+)}$ conversion is very similar to the 
$0\nu\beta\beta$-decay. Both processes violate lepton number 
by two units and take place only if the neutrino is a Majorana particle
with non-zero mass.
However, there are other important differences:
i) The available energies for these two processes differ considerably. 
In addition, the number of leptons in final states is different. These
facts result in significantly different phase space integrals.
ii) The emitted positron in $\mathbf{(\mu^-, e^+)}$ conversion 
has large momentum and therefore the long-wave approximation is not valid. 
iii) As it will be shown below, in the case of light neutrino-exchange 
there is a singular behavior
of the $\mathbf{(\mu^-, e^+)}$ nuclear matrix element which is the
additional source of the difficulties for the numerical integration.
iv) In the case of the $\mathbf{(\mu^-, e^+)}$ conversion 
there is a great number of 
nuclear final states. Nevertheless, the major contribution is here
assumed to come from  the transition
to the ground state of the final nucleus.  

\begin{figure}[t]
\centerline{\epsfig{file=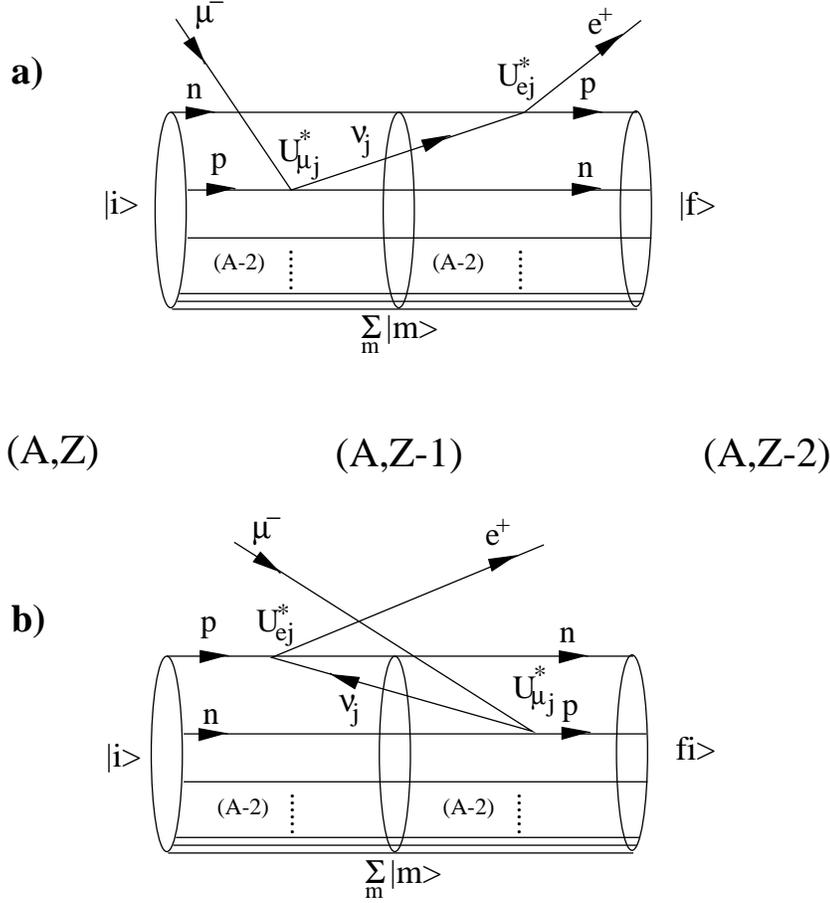,height=12.0cm}}
\caption{The direct (a) and cross (b) Feynmann diagrams of the 
 $(\mu^-,e^+)$ conversion in nuclei mediated by Majorana neutrinos.
}
\label{fig.1}
\end{figure}

We shall discuss the amplitude and width of $(\mu^-,e^+)$ conversion
in nuclei mediated by light Majorana neutrinos. This process is shown in 
Fig. \ref{fig.1}. We concentrate only on the 
nuclear transition connecting the ground states of the initial and final nuclei, 
which is favored 
from the experimental point of view 
due to the minimal background. 
In this case the $e^+$ spectrum has a peak at the energy 
\begin{equation} 
E_{e^+} = m_\mu - \varepsilon_b - (E_f - E_i).
\label{eq.12}
\end{equation} 
Here, $m_\mu$, $\varepsilon_b$, $E_i$ and $E_f$ are the mass of muon,
the muon atomic binding energy (for $^{48}Ti$ $\varepsilon_b = 1.45 ~MeV$),
the energies of initial and final ground states, respectively. 
Latter on we that the kinetic energy of the final nucleus is
negligible.

The weak 
interaction Hamiltonian in the neutrino mass eigenstate basis has the standard form
\begin{equation}
{\cal H}^{weak}(x)= U^{(\nu)}_{li}\frac{G_{{F}}}{\sqrt{2}} 
\left[\bar{l}_{{L}}(x)\gamma_\alpha (1 + \gamma_5) \nu_{{i L}}(x)\right]
j_\alpha(x) + {h.c.}~~~~ (l~ = ~ e, ~\mu, ~\tau),
\label{eq.13}
\end{equation}
where $j_{\alpha}(x)$ is the charged 
hadron 
current.
The neutrino mixing matrix $U^{(\nu)}_{li}$ is defined in Eq. (\ref{eq.8}). 

In second order of the weak-interaction 
we get for the $(\mu^-,e^+)$ conversion  the following matrix element
\begin{eqnarray}
<f|S^{(2)}|i> &=& i \left(\frac{G_F}{\sqrt{2}}\right)^2 
~<m_\nu >_{\mu e}~
\frac{1}{(2\pi)^{3/2}}\frac{1}{\sqrt{4 E_{\mu^-} E_{e^+}}} ~
u^T(k_{e^+}) C^{-1} (1+\gamma_5) u(k_{\mu^-})\times \nonumber \\
&& \frac{2 m_e m_\mu}{4\pi m_\mu {\cal R} } g^2_A 
~{\cal M}^{\Phi}_{<m_\nu >_{\mu e}} 
~2\pi \delta(E_{\mu^-} + E_i - E_f - E_{e^+}), 
\label{eq.14}
\end{eqnarray}
where 
\begin{equation}
{\cal M}^{\Phi}_{<m_\nu >_{\mu e}} = \frac{M^{\Phi}_{F}}{g^2_A}
- M^{\Phi}_{GT},
\label{eq.15}
\end{equation}
with
\begin{eqnarray}
M^{\Phi}_{F} &=& \frac{4 \pi {\cal R} }{(2 \pi)^3} 
\int \frac{d\vec{q}}{2 q} \times \nonumber \\
&& \sum_n \left(
\frac{ 
<0^+_i|\sum_l \tau^+_l e^{-i {\vec{k}}_{e^+} \cdot{\vec{r}}_l}
e^{-i \vec{q}\cdot{\vec{r}}_l} |n><n|
\sum_m \tau^+_m 
e^{i \vec{q}\cdot{\vec{r}}_m} \Phi (r_m) |0^+_f>}
{q - E_{\mu^-} + E_n - E_i + i \varepsilon_n}
 + \right. \nonumber \\
&& \left.
~~~\frac{ 
<0^+_i|\sum_m \tau^+_m 
e^{i \vec{q}\cdot{\vec{r}}_m} \Phi (r_m) 
|n><n|\sum_l \tau^+_l 
e^{-i {\vec{k}}_{e^+} \cdot{\vec{r}}_l}
e^{-i \vec{q}\cdot{\vec{r}}_l} 
|0^+_f>}
{q + E_{e^+} + E_n - E_i + i \varepsilon_n}
  \right),
\label{eq.16}
\end{eqnarray}
\begin{eqnarray}
M^{\Phi}_{GT} &=& \frac{4 \pi {\cal R} }{(2 \pi)^3} 
\int \frac{d\vec{q}}{2 q} \times \nonumber \\
&& \sum_n \left(
\frac{ 
<0^+_i|\sum_l \tau^+_l {\vec{\sigma}}_l 
e^{-i {\vec{k}}_{e^+} \cdot{\vec{r}}_l}
e^{-i \vec{q}\cdot{\vec{r}}_l} |n>\cdot <n|
\sum_m \tau^+_m {\vec{\sigma}}_m 
e^{i \vec{q}\cdot{\vec{r}}_m} \Phi (r_m) |0^+_f>}
{q - E_{\mu^-} + E_n - E_i + i \varepsilon_n}
 + \right. \nonumber \\
&& \left.
~~~\frac{ 
<0^+_i|\sum_m \tau^+_m {\vec{\sigma}}_m 
e^{i \vec{q}\cdot{\vec{r}}_m} \Phi (r_m) 
|n>\cdot <n|\sum_l \tau^+_l {\vec{\sigma}}_l 
e^{-i {\vec{k}}_{e^+} \cdot{\vec{r}}_l}
e^{-i \vec{q}\cdot{\vec{r}}_l} 
|0^+_f>}
{q + E_{e^+} + E_n - E_i + i \varepsilon_n}
 \right).
\label{eq.17}
\end{eqnarray}
Here, ${\cal R} = r_0 A^{1/3}$ 
is the mean nuclear radius, with $r_0 = 1.1 fm$
and $m_e$ is the  mass of  electron. 
${\vec r}_i$ is a coordinate of the $i$th nucleon.
$E_{\mu^-}$ ($k_{\mu^-}$) and $E_{e^+}$ ($k_{e^+}$) 
denote energies (four-momenta) of the bound muon 
($E_{\mu^-} = m_\mu - \varepsilon_b$) and the emitted
 positron, respectively.
$E_n$ and $\varepsilon_n$  are respectively energy and width of the 
intermediate nuclear state. $\Phi (r)$ is the radial part of  
bound muon in its orbit (see Appendix A). 

In the derivation of nuclear matrix element ${\cal M}^{\Phi}_{<m_\nu >_{\mu e}}$
we neglected the contribution from higher order terms of nucleon current 
(weak-magnetism, induced pseudoscalar coupling), which are expected to play
a less important role.  Following the analysis in Ref. 
\cite{pseu99} their consideration can reduce the value 
of ${\cal M}^{\Phi}_{<m_\nu >_{\mu e}}$ by an amount of about  $20\%$
by  analogy to the $0\nu\beta\beta$-decay.

We have normalized the  nuclear matrix element 
 ${\cal M}^{\Phi}_{<m_\nu >_{\mu e}}$ in the same way as 
usual for the corresponding $0\nu\beta\beta$-decay matrix
element. We note that the denominators in the expressions for
the $(\mu^-,e^+)$ conversion 
and the $0\nu\beta\beta$-decay 
exhibit a different behavior. 
This is because the energy of the bound
muon $E_{\mu^-}$ is large. The two denominators in Eq. (\ref{eq.17}) can be 
associated with the direct and the cross Feynman diagrams in Fig. \ref{fig.1}.
 One notes
that the value of $(- E_{\mu^-} + E_n - E_i)$ is negative. This fact 
implies that the widths of the nuclear states play an important role and 
that the imaginary part of the nuclear matrix element can be large. 
This point was not discussed in previous publications 
\cite{doi,kamal,vergr,leon} and is one of the motivations of our 
$(\mu^-,e^+)$ conversion calculation. 
We want to investigate if
this singular behavior of the amplitude can lead to an enhancement
of the $(\mu^-,e^+)$ conversion branching ratio or not.
In order to simplify the numerical calculations we complete the sum 
over virtual intermediate nuclear states 
by closure after replacing $E_n$, $\varepsilon_n$
by some average values
$<E_n>$, $\varepsilon$, respectively:
\begin{eqnarray}
\sum_n \frac{|n><n|}
{q - E_{\mu^-} + E_n - E_i + i \varepsilon_n}
~& = & ~ 
\frac{1}
{q - E_{\mu^-} + <E_n> - E_i + i \varepsilon}, \nonumber \\
\sum_n \frac{|n><n|}
{q + E_{e^+} + E_n - E_i + i \varepsilon_n}
~& = & ~
\frac{1}
{q + E_{e^+} + <E_n> - E_i + i \varepsilon} 
\label{eq.18}
\end{eqnarray}

Next we assume that the muon wave function
varies very little inside the nuclear system, i.e.,
the following approximation is used
\begin{equation}
|{\cal M}^{\Phi}_{<m_\nu >_{\mu e}}|^2 ~=
<\Phi_\mu >^2 ~|{\cal M}_{<m_\nu >_{\mu e}}|^2.
\label{eq.19}
\end{equation}
The explicit form of $<\Phi_\mu >^2$ is given in Appendix B.\\

For the width of $(\mu^-,e^+)$ conversion we obtain 
\begin{eqnarray}
\Gamma_{<m_\nu >_{\mu e}} = \frac{1}{\pi}~ E_{e^+}~ k_{e^+}~
F(Z-2,E_{e^+})~ 
c_{\mu e} ~
<\Phi_\mu >^2 ~
|{\cal M}_{<m_\nu >_{\mu e}}|^2 ~
\left|\frac{<m_\nu >_{\mu\ e}}{m_e}\right|^2,
\label{eq.20}
\end{eqnarray}
where $c_{\mu e}~ = 2 G_F^4[({m_e m_\mu})/{(4 \pi m_\mu {\cal R} )}]^2 g_A^4$
and $k_{e^+}~=~|{\vec k}_{e^+}|$. 
The nuclear matrix element ${\cal M}_{<m_\nu >_{\mu e}}$ 
can be decomposed 
into 
the contributions coming from direct and
crossed Feynmann 
diagrams
in Fig. \ref{fig.1} as
\begin{equation}
{\cal M}_{<m_\nu >_{\mu e}} = 
M^{dir.} +  M^{cro.},
\label{eq.21}
\end{equation}
where
\begin{eqnarray}
M^{dir.} &=& 
<0^+_i| ~\sum_{kl} \tau^+_k \tau^+_l
4\pi \sum_\lambda (-1)^\lambda \sqrt{2\lambda+1}
~j_\lambda (k_{e^+} R_{kl}) 
~\{ Y_\lambda (\Omega_r) \otimes Y_\lambda (\Omega_R) \}_0 \times \nonumber \\
&& ~~~~~ \frac{{\cal R}}{\pi} \int_0^\infty 
\frac{j_0(q r_{kl}) j_\lambda (k_{e^+} r_{kl}/2 )}
{q - E_{\mu^-} + E_n - E_i + i \varepsilon}
(\vec{\sigma_k}\cdot\vec{\sigma_l}f_A^2(q^2) -\frac{f_V^2(q^2)}{g^2_A})  
q dq ~|0^+_f> \nonumber \\
M^{cro.} &=& 
<0^+_i| ~\sum_{kl} \tau^+_k \tau^+_l
4\pi \sum_\lambda (-1)^\lambda \sqrt{2\lambda+1}
~j_\lambda (k_{e^+} R_{kl}) 
~\{ Y_\lambda (\Omega_r) \otimes Y_\lambda (\Omega_R) \}_0 \times \nonumber \\
&& ~~~~~\frac{{\cal R}}{\pi} \int_0^\infty 
\frac{j_0(q r_{kl}) j_\lambda (k_{e^+} r_{kl}/2 )}
{q + E_{e^+} + E_n - E_i + i \varepsilon}
(\vec{\sigma_k}\cdot\vec{\sigma_l}f_A^2(q^2) -\frac{f_V^2(q^2)}{g^2_A})  
q dq ~|0^+_f>,
\label{eq.22}
\end{eqnarray}
with
\begin{eqnarray}
\label{notation1}
{\vec r}_{ij}={\vec r}_i-{\vec r}_j, \ \ \
r_{ij}=|{\vec r}_{ij}|,\ \ \
{\vec R}_{ij}={\vec r}_i+{\vec r}_j, \ \ \
R_{ij}=|{\vec R}_{ij}|.
\ \ \
\label{eq.23}
\end{eqnarray}

For the normalized
nucleon form factors we use the conventional dipole form  
$f_V(q^2) = 1/(1+q^2/\Lambda^2_V)^2$ [$\Lambda^2_V = 0.71~(GeV)^2$],
$f_A(q^2) = 1/(1+q^2/\Lambda^2_A)^2$ [$\Lambda_A = 1.09~GeV$].

\section{Results and discussions}
\label{sec:level4}

The nuclear matrix elements of the $(\mu^-,e^+)$ conversion process
have been calculated within the proton-neutron renormalized
Quasiparticle Random Phase Approximation \cite{toi95,simn96,FKSS97,awf99}.
The considered single-particle model space both for protons and
neutrons have been as follows:
The full $0-3\hbar\omega$ shells
plus $2s_{1/2}$, $0g_{7/2}$ and $0g_{9/2}$ levels.  
The single particle energies were  obtained by using a  Coulomb--corrected 
Woods--Saxon potential. Two-body G-matrix elements
were calculated from the Bonn one-boson exchange potential 
within the Brueckner theory. 
The pairing interactions have been adjusted to fit 
the empirical pairing gaps \cite{cheo93}. 
The  particle-particle and particle-hole channels of 
the G-matrix interaction of the nuclear Hamiltonian $H$ 
are renormalized by introducing
the parameters $g_{pp}$ and $g_{ph}$, respectively. 
The calculations have been performed for $g_{ph} = 1.0$ and
$g_{ph} = 0.8,~1.0,~1.2$. The two-nucleon correlation effect 
has been considered in the same way as in Ref. \cite{pseu99,awf99}.

In calculation of the  $(\mu^-,e^+)$ conversion nuclear matrix elements 
we have used the fact that  the widths of the low lying nuclear
states are negligible in comparison with their energies. Therefore
we have carried out
the
calculation in the limit $\varepsilon \rightarrow 0$
using the formula
\begin{equation}
\frac{1}{\alpha + i\varepsilon} = 
{\cal{P}} \frac{1}{\alpha} - i \pi \delta(\alpha),
\label{eq.24}
\end{equation}
which allows
one to separate the real and imaginary 
parts
 of the 
$(\mu^-,e^+)$ conversion amplitude.

\begin{table}[t]
\caption{Nuclear matrix elements of the light Majorana neutrino exchange mode of
the $(\mu^-,e^+)$ conversion in $^{48}Ti$ [see Eqs. (\ref{eq.21}) 
and (\ref{eq.22})].
The calculations have been performed within pn-RQRPA 
without and with consideration
of two-nucleon short-range correlations (s.r.c.).}
\label{table.1}
\begin{tabular}{lccccccccc}\hline
 & \multicolumn{4}{c}{ without s.r.c} & 
\multicolumn{4}{c}{ with s.r.c} \\ \cline{2-5} \cline{7-10} 
 $g_{pp}$ &  
 $M^{cro.}$ & $R(M^{dir.})$ & $I(M^{dir.})$ & 
  $|{\cal M}_{<m_\nu >_{\mu e}}|$ & &
 $M^{cro.}$ & $R(M^{dir.})$ & $I(M^{dir.})$ & 
  $|{\cal M}_{<m_\nu >_{\mu e}}|$ \\
 & [$10^{-2}$] &[$10^{-2}$] &[$10^{-2}$] &[$10^{-2}$] & &
  [$10^{-2}$] &[$10^{-2}$] &[$10^{-2}$] &[$10^{-2}$] \\  \hline
 0.80 & 9.65 & 0.23 & 8.83 & 13.2 & & 
 4.88 & -7.99 & 4.98 & 5.87 \\
 1.00 & 7.71 & 3.36 & 5.88 & 12.5 & &
 3.40 & -4.03 & 2.37 & 2.45 \\
 1.20 & 5.05 & 9.09 & 1.78 & 14.2 & &
 1.30 &  2.71 & -1.32 & 4.22 \\ \hline
\end{tabular}
\end{table}
 
In Table \ref{table.1} nuclear matrix elements of the
light Majorana neutrino exchange mechanism of the $(\mu^-,e^+)$ conversion
in $^{48}Ti$ are presented.  
The adopted value of $<E_n>-E_i$ was 10 MeV.  We have found
that our
results depend weakly on this average value of
the nuclear states within the interval 
$2~MeV~ \le (<E_n>-E_i) \le ~15~MeV$. However, they
depend significantly on the details of
nuclear model, in particular, on the renormalization of the 
particle-particle channel of the nuclear Hamiltonian,
and on the two-nucleon short-range correlation effect (s.r.c.). 
A new feature of this $(\mu^-,e^+)$ conversion calculation is that 
the imaginary part of ${\cal M}_{<m_\nu >_{\mu e}}$
is significant, i.e., can not be neglected.
This fact was not noticed in the previous $(\mu^-,e^+)$ calculations
\cite{doi,kamal,vergr,leon}. 

In the further analysis we shall consider the nuclear matrix element
$|{\cal M}_{<m_\nu >^{\mu e}}|$ obtained for $g_{pp}=1.0$ by
considering the two-nucleon short range correlations. It is interesting to 
compare its value with the value of $0\nu\beta\beta$-decay matrix 
elements for A=48 nuclear system.  We have 
\begin{equation}
|{\cal M}_{<m_\nu >_{\mu e}}| = 2.45\times 10^{-2},~~~~~
|{\cal M}_{<m_\nu >_{ee}}| = 0.82.
\label{eq.25}
\end{equation}
We see that the matrix element for the $(\mu^-,e^+)$ conversion 
is strongly suppressed in comparison with $0\nu\beta\beta$-decay 
matrix element by about factor of 400. It is mostly due to the
large momentum of the outgoing positron in the $(\mu^-,e^+)$ conversion 
process. 

One can compare also the width of $(\mu^-,e^+)$ conversion in $^{48}Ti$
with the width of $0\nu\beta\beta$-decay of $^{48}Ca$. We get 
\begin{eqnarray}
\frac{\Gamma_{<m_\nu >_{\mu e^+}}}{\Gamma_{<m_\nu >_{ee}}} 
&=& \frac{ln(2)}{G_{01}}
\frac{1}{\pi}~ E_{e^+}~ K_{e^+}~
F(Z-2,E_{e^+})~ 
c_{\mu e} ~
<\Phi_\mu >^2 ~
\frac{|{\cal M}_{<m_\nu >_{\mu e}}|^2}{|{\cal M}_{<m_\nu >_{ee}}|^2}
 ~\left|\frac{<m_\nu >_{\mu e}}{<m_\nu >_{ee}}\right|^2,
\nonumber \\
&=& 1.97\times 10^5~\frac{|{\cal M}_{<m_\nu >_{\mu e}}|^2}{|{\cal M}_{<m_\nu >_{ee}}|^2}
 ~\left|\frac{<m_\nu >_{\mu e}}{<m_\nu >_{ee}}\right|^2, \nonumber \\
&=& 176.
 ~\left|\frac{<m_\nu >_{\mu e}}{<m_\nu >_{ee}}\right|^2.
\label{eq.26}
\end{eqnarray} 
The width of the $(\mu^-,e^+)$ 
conversion is enhanced mostly due to the larger available energy for
this process. A comparison with the width of 
$0\nu\beta\beta$-decay show that it is disfavored by smaller coulombic
factor F(Z,E) ($\sim 0.623/1.8$) and by significantly  smaller
value of associated nuclear matrix element 
($\sim (0.0245/0.82)^2$). If we assume the effective neutrino masses
${<m_\nu >_{\mu e}}$ and  ${<m_\nu >_{ee}}$ to be comparable 
[see Eq. (\ref{eq.11})] we find
that ${\Gamma_{<m_\nu >_{\mu e^+}}}$ is enhanced by a factor of 
about 200 in comparison with ${\Gamma_{<m_\nu >_{ee}}}$. 
We have used   $G_{01} = 8.031\times 10^{-14}~year^{-1}$ 
\cite{pantis}.

From the experimental point of view 
it is interesting to compare the $(\mu^-,e^+)$ conversion width with the
width of ordinary muon capture rate. We have 
\begin{eqnarray}
\frac{\Gamma_{<m_\nu >_{\mu e}}}{\Gamma_\mu} &=& 
2 \frac{E_{e^+}~ k_{e^+}}{m^2_\mu} 
\frac{c_{\mu\mu}}{G_F^2} 
\frac{|{\cal M}_{<m_\nu >_{\mu e}}|^2 F(Z-2,E_{e^+}) }
{[G^2_V + 3 G^2_A + G_P^2 - 2 G_A G_P] Z f(Z,A)}
\left|\frac{<m_\nu >_{\mu e}}{m_e}\right|^2 \nonumber \\
&=& 2.24\times 10^{-22} 
|{\cal M}_{<m_\nu >_{\mu e}}|^2 
\left|\frac{<m_\nu >_{\mu e}}{m_e}\right|^2 \nonumber \\
&=& 1.34\times 10^{-25} 
\left|\frac{<m_\nu >_{\mu e}}{m_e}\right|^2 
\label{eq.27}
\end{eqnarray} 
If we use the prediction for $<m_\nu >_{\mu e}$ coming
from neutrino oscillation phenomenology, i.e., 
$<m_\nu >_{\mu e}~ \le~ 0.97 ~eV $ and more conservative bound
$<m_\nu >_{\mu e}~ \le~ 9 ~eV $
 [see Eq. (\ref{eq.11})],  we end up
with
\begin{eqnarray}
\frac{\Gamma_{<m_\nu >_{\mu e}}}{\Gamma_\mu} = 
4.8\times 10^{-37},\ 4.5 \times 10^{-35},
\label{eq.28}
\end{eqnarray} 
This value is about ten orders of magnitude smaller as the 
estimated one for the $(\mu^-,e^+)$ conversion in $^{32}S$ by
Doi et al. \cite{doi}. 
There could be various reasons for this difference.
First,  the $<m_\nu >_{\mu e}$ nuclear matrix element
calculated by Doi et al. contains contributions from 
all final nuclear states and not only from the 
$0^+_{g.s.} \rightarrow 0^+_{g.s.}$ transition as in our case. 
It can be that 
the experimentally interesting $g.s. \rightarrow g.s.$ transition
exhausts only a small part from all allowed nuclear transitions.
Second, in the simplified calculation of Doi et al. nuclear
matrix elements were evaluated by summing the final nuclear 
states with closure. It usually leads to overestimation 
of the results as we know from calculation of ordinary 
muon capture. Third, the nuclear matrix element of Ref. \cite{doi}
have been evaluated by using the long--wave approximation. Our 
comparison of the $(\mu^-,e^+)$ conversion and the $0\nu\beta\beta$-decay 
(long--wave approximation is used)
matrix elements shows that it can lead to overestimation
of ${\cal M}_{<m_\nu >_{\mu e}}$  by factor up to $10^2$.
Fourth, the problem of the ground and short--range correlations have been
not addressed in Ref. \cite{doi}. We have found that  
 ${\cal M}_{<m_\nu >_{\mu e^+}}$
matrix element for A=48 nuclear system is strongly suppressed
by both of them. It is not clear whether this effect is
due to the chosen target. However, we note that ${\cal M}_{<m_\nu >_{\mu e^+}}$ 
for the conversion in $^{48}Ti$ consists of transition to the doubly closed
shell nucleus $^{48}Ca$, which, e.g., in the case of
$0\nu\beta\beta$-decay are less favored. To clarify this issue,
the  calculations of the  $(\mu^-,e^+)$ conversion for other nuclear
systems are necessary.  

It is worthwhile to notice that our result is in relative good agreement 
with the calculations performed by Leontaris and Vergados \cite{leon} for 
the $(\mu^-,e^+)$ conversion in $^{58}Ni$. By using the same value
for effective Majorana neutrino mass $<m_\nu >_{\mu e}$ as in this article
the result of Ref. \cite{leon} corresponds to a branching ratio equal to
$3.2\times 10^{-36}$ relative to the total absorption of the muon 
for the ground state to ground state transition.
The difference of about one order with our result for A=48 can be attributed
to the nuclear physics aspect of the  $(\mu^-,e^+)$ conversion, i.e.,
to a given nuclear system and the chosen nuclear model. We also remark that 
in Ref. \cite{leon} the imaginary part of the  $(\mu^-,e^+)$ conversion
amplitude was not considered. 

\section{Summary and outlook}
\label{sec:level5}

In summary, the lepton number violating process of $(\mu^-,e^+)$ conversion
in nuclei have been studied. The light Majorana neutrino--exchange
mechanism of this process have been considered. A detailed analysis 
of this mode of the $(\mu^-,e^+)$ conversion 
have been performed. The first realistic calculation of the 
$0^+_{g.s.} \rightarrow 0^+_{g.s.}$ channel of this process,
which is most favored for experimentally studies, are presented.
The relevant matrix elements for A=48 nuclear system have been 
calculated within the pn-RQRPA. The effects of the ground state 
and two-nucleon short-range correlations have been analyzed. 
It was found that by inclusion of them the value of  $(\mu^-,e^+)$
conversion matrix elements is strongly suppressed. We are the
first, to our knowledge, showing that the imaginary part of the nuclear matrix
element is large and should be taken into account. Further,
a comparison of different relevant aspects with the $0\nu\beta\beta$-decay  
process are presented. It is shown that the width of 
 $(\mu^-,e^+)$ conversion is about by factor of 200 larger as that
of  $0\nu\beta\beta$-decay by assuming predictions for effective
neutrino masses coming from neutrino oscillation phenomenology.
Nevertheless, the studied neutrino exchange mode of lepton number 
violating $(\mu^-,e^+)$ conversion is not suitable for experimental 
study being extremely small compared to the ordinary muon capture. This fact, 
however, does not disfavor further experimental study of the $(\mu^-,e^+)$ conversion
in nuclei as some other lepton number violating mechanisms, e.g., those
coming from GUT's and SUSY models,  can dominate this process. Therefore,
they should be carefully examine too. 

\vskip10mm
\centerline{\bf Acknowledgments}
We are grateful to I. Schmidt for useful comments and remarks. 
This work was supported in part by Fondecyt (Chile) under 
grant 8000017 and by RFBR (Russia) under grant 00-02-17587. 
\bigskip

\section{Appendix A}
\label{sec: levelA}

{The bound muon wave function (1S-state)} is  
\begin{equation}
\Psi_\mu (x) = \Phi_\mu (\vec{x}) e^{- i E_{\mu^-} x_0} 
\frac{u_\mu^s}{\sqrt{2 E_{\mu^-}}},
\end{equation}
where
\begin{eqnarray}
\Phi_\mu (\vec{x}) &=& \frac{Z^{3/2}}{(\pi a_\mu^3)^{1/2}}
e^{- Z |\vec{x}|/a_\mu}, \nonumber \\
u^s_\mu &= &
\left(\begin{array}{c} \chi^s\\
0 \end{array}\right) \sqrt{2 E_{\mu^-}}
\end{eqnarray}
with $a_\mu = 4\pi/(m_\mu e^2)$ 
($a_\mu/a_e \approx m_e/m_\mu \approx 5\times 10^{-3}$)
 $m_\mu$ is the reduced mass of muon
nucleus system.  

\section{Appendix B}
\label{sec: levelB}

The width for the ordinary muon capture reaction
$\mu^- + (Z,A) \rightarrow \nu_\mu + (Z-1,A)$ can be written 
in the Primakoff form \cite{pri}
\begin{eqnarray}
\Gamma_\mu = \frac{1}{2\pi} m^2_\mu  (G_F \cos \theta_c)^2 <\Phi_\mu >^2 Z 
[G^2_V + 3 G^2_A + G_P^2 - 2 G_A G_P] f(Z,A),
\end{eqnarray}
where muon average probability density over the nucleus is
\begin{equation}
<\Phi_\mu >^2 \equiv \frac{\int |\Phi_\mu (\vec{x})|^2 \rho(\vec{x}) d^3 x}
{\int \rho(\vec{x}) d^3 x}.
\end{equation}
$\rho(\vec{x})$ is the nuclear density. To a good approximation 
it has been found 
\begin{equation}
<\Phi_\mu >^2 = \frac{ \alpha^3 m_\mu^3}{\pi} \frac{Z^4_{eff}}{Z},
\end{equation}
i.e., the deviation from the behavior of the wave function at the 
origin  has been taken into account by the effective proton number 
$Z_{eff}$. The values of this effective charge has been calculated 
for the nuclear systems of interest in Ref. \cite{ko94}. In particular,
one finds $Z_{eff} = 17.6$ for $Z=22$. The quadratic combination of 
the weak coupling constants is
\begin{equation}
[G^2_V + 3 G^2_A + G_P^2 - 2 G_A G_P] \approx 5.9.
\end{equation}
The function $f(Z,A)$ takes into account the two-nucleon correlations
given by \cite{leon}
\begin{equation}
f(A,Z) = 1 - 0.03 \frac{A}{2 Z} + 0.25 (\frac{A}{2 Z} - 1 ) +
3.24 ( \frac{Z}{2 A} - \frac{1}{2} - |\frac{1}{8 Z} - \frac{1}{4 A}| ).
\end{equation}
This Pauli blocking factor for $^{48}Ti$ takes the value 
$f(22,48) = 0.11$.


\begin{thebibliography}{99}
%
\bibitem{theorem} J. Schechter and J.W.F. Valle, Phys.Rev. D { 25} 
(1982) 2951.   
\bibitem{DGKS:NANPino} C. Dib, V. Gribanov, S. Kovalenko and I. Schmidt,
to appear in the Proceedings of NANPino conference, Dubna, Russia,  
July 19-22, 2000; hep-ph/hep-ph/0011213.  
\bibitem{Mohapatra:mu} J.H. Missimer, R.N. Mohapatra and N.C. Mukhopadhyay,
                { Phys. Rev.}  { D 50} (1994) 2067. 
%
\bibitem{znbb-exp} L. Baudis et al., {Phys.Rev.Lett.} {\bf 83} (1999) 411. 
\bibitem{ejiri} H. Ejiri, Phys. Rep. 338 (2000) 265.
\bibitem{znbb-rev2} H.V. Klapdor-Kleingrothaus,
         {\it Springer Tracts in Modern Physics} {\bf 163} (2000) 69.
\bibitem{doi} M. Doi, T. Kotani, E. Takasugi, Prog. Theor. Phys. Suppl. 83
 (1985) 1.                                                                           
\bibitem{fae98} A. Faessler and F. \v Simkovic, J. Phys. { G} 24 (1998) 2139.
\bibitem{kamal} A.N. Kamal and J.N. Ng, Phys. Rev. D 20 (1979) 2269.
\bibitem{vergr} J.D. Vergados, Phys. Rev. D 23 (1981) 703, 
  Phys. Rep. 133 (1986) 1.
\bibitem{leon} G.K. Leontaris and J.D. Vergados, Nucl. Phys. B 224 (1983) 137.
\bibitem{ko94}  T.S. Kosmas, G.K. Leontaris and J.D. Vergados,
     Prog. Part. Nucl. Phys. { 33} (1994) 397.
\bibitem{ko97} T.S. Kosmas, Amand Faessler and J.D. Vergados,
       J. Phys. { G 23} (1997) 693.
\bibitem{mue} A. Faessler, T.S. Kosmas, S.G. Kovalenko, J.D. Vergados,
        Nucl. Phys. B 587 (2000) 25.
\bibitem{siis} T. Siiskonen, J. Suhonen, T.S. Kosmas, Phys. Rev. C 
    62 (2000) 035502.
\bibitem{doh93} C. Dohmen et al, SINDRUM II Collaboration, Phys. Lett. 
    { B 317} (1993) 631.
\bibitem{hon96} W. Honecker et al, SINDRUM II Collaboration, Phys. Rev. 
    Lett. { 76} (1996) 200.
\bibitem{toi95} J. Toivanen  and J. Suhonen, 
   Phys. Rev. Lett. 75 (1995) 410.
\bibitem{simn96} J. Schwieger, F. \v Simkovic, and A. Faessler,
   Nucl. Phys. { A 600} (1996) 179.
\bibitem{bil87} S.M. Bilenky and S.T. Petcov, Rev. Mod. Phys. 59 (1987) 671.
\bibitem{bil98} S.M. Bilenky, C. Giunti, W. Grimus, 
   Prog. Part. Nucl. Phys. { 43} (1999) 1.
\bibitem{zubmas} K. Zuber, Phys. Rep. { 305} (1998)  295.
\bibitem{SK-global} Super-Kamiokande Collab., S. Fukuda {\it et al.}, 
Phys. Rev. Lett. 85 (2000) 3999.
\bibitem{mix99} S.M. Bilenky, C. Guinty, W. Grimus, B. Kayser, and
    S.T. Petcov, Phys. Lett. B 465 (1999) 193.
\bibitem{viss} F. Vissani, JHEP 9906 (1999) 22. 
\bibitem{barg} V. Barger and K. Whisnant, Phys. Lett. B 456 (1999) 194.
\bibitem{smi} A.Yu. Smirnov,  Nucl. Phys. Proc. Suppl. 87 (2000) 288.
\bibitem{Lnv-oscil}  W. Rodejohann, hep-ph/0003149; hep-ph/0008044;
H.V. Klapdor-Kleingrothaus, H. P\"as and A.Yu. Smirnov, hep-ph/0003219
and references therein.
\bibitem{pseu99} F. \v Simkovic, G. Pantis, J.D. Vergados, and
    A. Faessler, Phys. Rev. C 60 (1999) 055502.
\bibitem{nanp}
    A. Faessler, F. \v Simkovic, Phys. At. Nucl. 63 (2000) 1165,
    hep-ph/9909252.
%
%
%
%
\bibitem{LS:2000}  L.S. Littenberg and R.E. Shrock, hep-ph/0005285.
\bibitem{grib} C. Dib, V. Gribanov, S. Kovalenko, I. Schmidt, 
    Phys. Lett. B 493 (2000) 82.
\bibitem{zuber} K. Zuber, Phys. Lett. B 479 (2000) 33.    
%
%
\bibitem{haug} O. Haug, J.D. Vergados, Amand Faessler, S. Kovalenko,
    Nucl. Phys. B 565 (2000) 38.
\bibitem{barg1} F. Vissani, hep-ph/9708483; V. Barger, T.J. Weiler, 
 and K. Whisnant, {Phys. Lett.} { B 442} (1998) 255.    
\bibitem{FKSS97} A. Faessler, S. Kovalenko, F. \v Simkovic,
   and J. Schwieger,
   Phys. Rev. Lett. { 78} (1997) 183  (1997);
  A. Faessler, S. Kovalenko, and F. \v Simkovic,
   Phys. Rev. { D} { 58} (1998) 115004.
\bibitem{awf99} A. Wodecki, W.A. Kami\'nski, and F. \v Simkovic,
   Phys. Rev. { D} 60 (1999) 11507. 
\bibitem{cheo93} M.K. Cheoun, Amand Faessler, F. \v Simkovic, G. Teneva,
         A. Bobyk,  Nucl.Phys. { A 561} (1993) 74.
\bibitem{pantis} G. Pantis, F. Simkovic, J.D. Vergados, A. Faessler,
    Phys. Rev. C 53 (1996) 695.
\bibitem{pri} H. Primakoff, Rev. Mod. Phys. 31 (1959) 802;
B. Goulard and H. Primakoff, Phys. Rev. C 11 (1975) 1894. 
\end{thebibliography}
\end{document}